\documentclass[11pt]{article}
\usepackage{amssymb,amsfonts,euscript,graphics,graphicx}
\usepackage{fancyhdr}

\begin{document}

\title{Efectos relativistas en los sistemas Galileo, GPS y GLONASS}
\vspace{0.6 cm}

\author{{\sc J.-F. Pascual-S\'anchez} \\[0.3cm]
{\sl Universidad de Valladolid, Espa\~na, UE } \\
{\sl e-mail: jfpascua@maf.uva.es}}

\date{}
\maketitle
\begin{abstract}
Actualmente, los Sistemas de Navegaci\'on Global por Sat\'elites
(GNSS), en funcionamiento s\'olo como sistemas de posicionamiento
global, son el  GPS (NAVSTAR) y el GLONASS, los cuales
\'unicamente son  operativos cuando se corrigen varios efectos
relativistas. En los pr\'oximos a\~nos se construir\'a el sistema
Galileo,  copiando el sistema GPS si no hay un proyecto
alternativo. En este trabajo se expondr\'a que hay una alternativa
a la mera copia, mediante el proyecto SYPOR, usando conceptos
relativistas y sin utilizar las ideas newtonianas que est\'an en
la base, tanto del GPS como del  GLONASS. Seg\'un el proyecto
SYPOR, el sistema Galileo ser\'{\i}a exacto, sin necesidad de
correcciones, y tendr\'{\i}a ventajas tecnol\'ogicas adicionales.

\end{abstract}

\section{Introducci\'on}
 En la actualidad, todos los sistemas de navegaci\'on global por
sat\'elite (GNSS), en funcionamiento s\'olo como sistemas de
posicionamiento global, es decir el GPS (NAVSTAR) (USA) y el
GLONASS (Rusia), est\'an basados en un modelo Newtoniano que usa
como sistema de referencia para la navegaci\'on un sistema no
inercial que gira con la tierra y centrado en ella, ECEF (WGS-84),
y como sistema de referencia temporal un tiempo coordenado de un
sistema de referencia inercial centrado en la Tierra, ECI,
corregido num\'ericamente por numerosos e importantes ``efectos
relativistas". Entre ellos, los m\'as significados son: el
desplazamiento gravitacional de  la frecuencia del reloj del
sat\'elite (Relatividad General) y el desplazamiento Doppler de
segundo orden, debido al movimiento del sat\'elite (Relatividad
Especial). Si no fueran corregidos, el sistema ser\'{\i}a
err\'oneo a partir de unos minutos. Durante un d\'{\i}a,
supondr\'{\i}an un error en el posicionamiento horizontal  del
receptor de m\'as de 11 Kil\'ometros, durante una semana, el error
acumulado en el posicionamiento vertical ser\'{\i}a de unos 5 Km.

Sin embargo, una constelaci\'on de sat\'elites cuyos relojes
intercambian tiempos propios mediante se\~nales
electromagn\'eticas es en realidad un sistema relativista o
Einsteiniano. Seg\'un el proyecto SYPOR (SIstema de
POsicionamiento Relativista) para el Galileo (UE), el segmento de
control deber\'{\i}a estar en los sat\'elites y por tanto
coincidir con el segmento espacial y no en el geoide terrestre
como en los sistemas actuales. De  esta manera, se invertir\'{\i}a
la lectura usual y la funci\'on del segmento de control no
ser\'{\i}a determinar la posici\'on de los sat\'elites con
respecto a ciertas coordenadas terrestres, sino definir \'estas
con respecto a la constelaci\'on de sat\'elites.

Seg\'un el proyecto SYPOR, el sistema Galileo  ser\'{\i}a exacto y
sin necesidad de correcciones. Globalmente, la situaci\'on actual
en los GNSS es an\'aloga a la siguiente: imag\'{\i}nese que un
siglo despu\'es de Kepler (Einstein), los astr\'onomos continuaran
usando las leyes de Kepler (Einstein) como un algoritmo para
corregir los epiciclos Ptolemaicos (Newton) por medio de ``efectos
Keplerianos (relativistas o Einsteinianos)".

 \section{Caracter\'{\i}sticas principales del GPS}
 El sistema GPS est\'a constitu\'{\i}do por tres partes o segmentos
 diferenciados:


\noindent {\bf 1) Segmento espacial}: El Sistema del
Posicionamiento Global (GPS) est\'a compuesto por una
constelaci\'on de 24 sat\'elites distribu\'{\i}da en 6 planos
orbitales igualmente espaciados de 55 grados de inclinaci\'on con
respecto al plano equatorial. Cada plano contiene 4 sat\'elites
(SV) a una altitud media de 20,183 Km, en \'orbitas de baja
excentricidad (la m\'as alta, la del SV nº13 es e\,=\,0.01486),
con periodo orbital de 12 horas sid\'ereas. Por ello, la misma
configuraci\'on aparece cada d\'{\i}a a la misma hora menos 4
minutos. Un receptor fijo en la Tierra observar\'a a un sat\'elite
dado dos veces al d\'{\i}a y habr\'a al menos cuatro sat\'elites
observables desde la Tierra en cualquier momento. En cada
sat\'elite  opera un reloj at\'omico de cesio (precisi\'on interna
de 4 nanosegundos por d\'{\i}a) y emite se\~nales
electromagn\'eticas que incluyen un c\'odigo que se\~nala al
receptor su posici\'on con respecto a la superficie de la Tierra
en un sistema de referencia que gira con la Tierra y centrado en
ella ECEF, (WGS-84), y su tiempo coordenado GPS (adelantado sobre
el UTC).\\
 {\bf 2) Segmento de control}: consta b\'asicamente de una
estaci\'on principal situada en Colorado Springs (la cual
transmite correcciones a las efem\'erides de los sat\'elites y a
sus relojes, usando UTC) y otras cuatro estaciones
cuasiecuatoriales en Hawai, Kwjalein, Ascensi\'on y Diego
Garc\'{\i}a.\\
 {\bf 3) Segmento de usuario}: Analizando se\~nales al menos
4 de estos sat\'elites, un receptor en la superficie de Tierra
puede conocer
 su posici\'on (latitud, longitud, y altitud)
 con una exactitud de al menos 5 metros, aunque con t\'ecnicas
 denominadas diferenciales, el error pueda reducirse a algunos cent\'{\i}metros.

\section{Geodesia cuadridimensional}

 El problema de navegaci\'on para un receptor en la
Tierra (T) consiste en computar su posici\'on espacio-temporal
$({\mathbf R}_T, {\mathbf T}_T)$, a partir de las cuatro
posiciones espacio-temporales emitidas por cuatro sat\'elites
$({\mathbf r}_{S{_j}}, t_S{_j}), $ $j\,=\, 1,2,3,4$.

Las ecuaciones de navegaci\'on del GPS est\'an basadas en la
aplicaci\'on del
 {\it segundo principio de
Relatividad Especial} (constancia de la velocidad de la luz en el
vac\'{\i}o), en  coordenadas usuales Minkowskianas, son:\\
\begin{equation}\label{2}
|{\bf R} _T \, -\,{\bf r}_{S{_j}} |^2 =
c^2\,(T_T\,-\,t_{S{_j}})^2\; ; \,j=1,2,3,4\,.
\end{equation}
Estas son las ecuaciones de los cuatro rayos de luz (se\~nales
electromagn\'eticas) que viajan sobre los {\it 4 ``conos de luz"
}(hipersuperficies 3-dim) desde los sat\'elites al receptor en el
espacio-tiempo de Minkowski (sin gravedad) y que, por lo tanto,
{\it solo son v\'alidas en un sistema de referencia inercial}. En
el GPS se considera que \'este es el sistema newtoniano de {\it
referencia inercial centrado en la Tierra} (ECI) y que no gira con
ella al apuntar a estrellas (en realidad, cu\'asares) fijas.

Desde el punto de vista de los sistemas de referencia newtonianos,
la primera misi\'on del GPS es determinar la posici\'on
tridimensional del receptor en la Tierra (este-oeste, norte-sur y
vertical) en el sistema ECEF. En principio, las se\~nales de tres
sat\'elites proporcionan esta informaci\'on. Cada sat\'elite
env\'{\i}a un se\~nal que codifica donde est\'a el sat\'elite  y
el tiempo de emisi\'on del se\~nal. El  reloj del receptor
cronometra la recepci\'on de cada se\~nal, entonces substrae el
tiempo de la emisi\'on, para determinar el lapso de tiempo (de 67
ms a 86 ms)  y la distancia que la se\~nal ha viajado (a la
velocidad de luz). \'Esta era la distancia  a la que el sat\'elite
estaba cuando emiti\'o la se\~nal. Tomando estas tres distancias
como radios, se construyen tres esferas  con centro en cada
sat\'elite. El receptor est\'a localizado en uno de los dos puntos
de intersecci\'on de las tres esferas. Este es, en resumen, el
procedimiento de trilateraci\'on.

Sin embargo, todav\'{\i}a hay un problema: el reloj del receptor
es de cuarzo y no es tan exacto como los relojes at\'omicos de los
sat\'elites. Por esta raz\'on, se necesita la se\~nal de un cuarto
sat\'elite para obtener el error en la precisi\'on del reloj del
receptor y habilita a \'este como si fuera un reloj at\'omico.

Veremos posteriormente que desde el punto de vista de los sistemas
de posicionamiento relativistas, el receptor determina sus 4
coordenadas (en general como veremos, no necesariamente
.
espaciales y 1 temporal, sino cuatro tiempos propios),
identificando los puntos de corte de los 4 conos de luz de los
sat\'elites que interseccionan a su cono de luz
propio.\\

\section{Errores Relativistas en los GNSS}
La precisi\'on temporal requerida por el GPS es del orden de $1\,
{\rm ns}\, = 10^{-9} \,{\rm seg}\,$, por lo cual, los errores
introducidos por los efectos relativistas fueron centrales en su
concepci\'on. El olvido de estos efectos har\'{\i}an al GPS
in\'util por tres razones principales.

{\bf 1.-Primero: efecto gravitatorio Einstein}.
 Porque los relojes se adelantan
(desplazamiento hacia el violeta) cuando est\'an m\'as alejados de
un centro de atracci\'on gravitatoria.

{\bf 2.-Segundo: efecto cinem\'atico Doppler relativista}.
 El movimiento del sat\'elite debe tenerse en cuenta, ya
que da lugar al efecto Doppler cl\'asico (lineal en velocidad) y
al efecto relativista especial Doppler de segundo orden en
velocidad (los relojes en movimiento se atrasan).

{\bf 3.-Tercero: efecto Sagnac y de campo gravitom\'agnetico de
rotaci\'on}.
 En principio, la rotaci\'on de la Tierra deber\'{\i}a
tenerse en cuenta por dos razones. En primer lugar, por el efecto
cinem\'atico Sagnac (ver por ej. \cite{pas}) y en segundo lugar,
por el efecto del campo gravitomagn\'etico (GM) (ver por ej.
\cite{mas}),
 generado por la rotaci\'on de la masa terrestre (hasta
ahora los efectos de este campo GM no han sido incorporados a los
GNSS).

{\it Para corregir los efectos relativistas, en el GPS se utiliza
la m\'etrica de Schwarzschild para el espacio-tiempo cercano a la
Tierra}. Sin embargo, estrictamente, la m\'etrica est\'atica y de
simetr\'{\i}a esf\'erica de Schwarzschild {\it no} describe
exactamente el espacio-tiempo de la Tierra por dos razones:

  1) La Tierra no es  esf\'erica.

  2) La Tierra  no permanece est\'atica sino que, al estar
girando, genera un campo de gravedad de rotaci\'on (no newtoniano)
o gravitomagn\'etico
 (GM) y debe ser
descrita por una {\it m\'etrica de Kerr} (o su aproximacion lineal
Lense-Thirring). Pero la Tierra gira despacio,
 $\omega_{_{T}} =  7.2921151247 \times  10^{-5} \;{\rm rad}\; s^{-1}$ ($v _{_{T}}= 465 \;{\rm m}/s$
  en el ecuador) y {\it la m\'etrica de
Schwarzschild es una  aproximaci\'on suficiente, por ahora}, con
la precisi\'on  en la medida del tiempo (ns), para los
prop\'ositos de los GNSS.

 N\'otese sin embargo, que seg\'un las resoluciones
de la IAU-2000, los efectos gravitom\'agneticos deben ser
introducidos para definir, tanto el  sistema de referencia celeste
baric\'entrico (BCRS), como el  sistema de referencia celeste
geoc\'entrico (GRCS).

 {\it La expresi\'on de la m\'etrica de Schwarzschild}, considerando
 \'orbitas equatoriales, es:
\begin{equation}\label{1}
ds^2 =  \left(1-\frac{2GM}{rc^2}\right)\,c^2\,dt^2 - \left
(1-\frac{2GM}{ r\,c^2}\right)^{-1} (dr ^2 + r^2 d\phi^2)~,
\end{equation}
donde $s$ es tiempo propio, {\it $t$ es un tiempo coordenado
medido por un observador inercial situado en el infinito} y $\phi$
es el \'angulo azimutal.

Apliquemos esta ecuaci\'on
  al reloj del
sat\'elite y a un reloj fijo en el  ecuador de Tierra y girando
con ella. Ambos, el reloj de la Tierra y del sat\'elite viajan a
distancia constante alrededor del centro de Tierra, por lo tanto
$dr = 0$ para cada reloj. Para ambos se obtiene:

\begin{equation}\label{3}
\left(\frac{ds}{c\,dt}\right)^2 =  \left(1+\frac{2\,V}{ c^2
}\right)\,-\,\frac{v^2}{c^2}\,.
\end{equation}

Donde $V$ es el potencial Newtoniano y
   $v = r \frac{d\phi}{dt}$ es la
velocidad tangencial a lo largo de la \'orbita circular
equatorial, medida usando el tiempo coordenado en el infinito $t$.

Si aplicamos dos veces la ecuaci\'on anterior, primero al
sat\'elite (S), (utilizando $r = r_{_{S}}$ , $v = v_{_{S}}$ y
tiempo propio $ds = c\,dt_{_{S}}$) y
 segundo, a la Tierra (T), (utilizando, $r
= r_{_{T}}$ , $v = v _{_{T}}$ y tiempo propio $ds =
c\,dt_{_{T}}$), con el mismo lapso de tiempo, $dt$, del observador
inercial en el infinito, se obtiene, dividiendo ambas expresiones:

\begin{equation}\label{4}
\left(\frac{dt_{T}}{dt_{S}}\right)^2=
\frac{1-\displaystyle\frac{2\,GM}{r_{_{T}}\,c^2}\,-\,\frac{v^2_{_{T}}}{c^2}
}{1-\displaystyle\frac{2\,GM}{r_{_{S}}\,c^2}\,-\,\frac{v^2_{_{S}}}{c^2}}.
\end{equation}

Los efectos relativistas que estamos considerando son,
\'unicamente, los del {\it orden $O(c^{-2})$.

 Este es el orden de aproximaci\'on que se usa en el GPS.}

 ?`Cu\'ando esta  aproximaci\'on es suficiente? Cuando el error
 intr\'{\i}nseco
 es de $4\, ns/{\rm dia}$, que corresponde a una desviaci\'on de Allan
  de  $5\times10^{-14}$ en los relojes at\'omicos del GPS.
 {\it ?`Los efectos relativistas son peque\~nos a este orden? No}, como
 veremos son m\'as que importantes, son cruciales.

 Hay que se\~nalar por otra parte, que la consideraci\'on de otros efectos relativistas
 al orden $O(c^{-3})$ son necesarios cuando se utilizan relojes at\'omicos
  enfriados por laser de \'ultima generaci\'on, con errores medibles  del
  orden $ps= 10^{-3}\,ns$ como en la misi\'on ACES (Atomic Clock Ensemble in Space)
   de la ESA para la ISS,
 y los de orden  $O(c^{-4})$, cuando se
 trabaje en el pr\'oximo futuro con errores de $fs:= 10^{-3}\,ps$.

 \subsection{Efecto gravitatorio Einstein:\,45,700 \,\,ns/d\'{\i}a}
  Comenzamos ignorando los movimientos de reloj
del sat\'elite y del reloj en la superficie de Tierra, i.e., el
efecto Doppler de segundo orden de Relatividad Especial,
suponiendo que ambos est\'an fijos en el ECI: $v _{_{T}}=
v_{_{S}}=  0$.
 Entonces, la ecuaci\'on (\ref{4}) queda reducida a
 \begin{equation}\label{6}
 \frac{dt_{T}}{dt_{S}}=
\frac{\left(1-\displaystyle\frac{2\,GM}{r_{_{T}}\,c^2}
\right)^\frac{1}{2}}{\left(1-\displaystyle\frac{2\,GM}{r_{_{S}}\,c^2}\right)^\frac{1}{2}}\,.
 \end{equation}

 El radio de una  \'orbita circular  con un per\'{\i}odo de  12 horas sid\'ereas
 es aproximadamente $r_{_{S}}=
 26.6\times10^{8} $ cm.
 Si se utiliza la aproximaci\'on lineal del binomio de Newton
 \begin{equation}\label{7}
(1\,+\,d)^{n} \thickapprox 1 \,+ n\, d\,,
 \end{equation}
ya que, por ejemplo,
\begin{equation}\label{8}
d= \displaystyle\frac{GM}{r_{_{T}}\,c^2}= 6.9\times10^{-10}\ll{1}
\end{equation}
y
\begin{equation}\label{9}
d= \displaystyle\frac{GM}{r_{_{S}}\,c^2}=
1.6\times10^{-10}\ll{1}\,,
\end{equation}
se obtiene
\begin{equation}\label{10}
\frac{dt_{T}}{dt_{S}}= 1-\displaystyle\frac{GM}{r_{_{T}}\,c^2}
+\frac{\,GM}{r_{_{S}}\,c^2} =1- \,D\,.
\end{equation}
 El n\'umero positivo representado por $D$ en la
ecuaci\'on (\ref{10}), es una estimaci\'on de la dilataci\'on
temporal entre los relojes estacionarios en la posici\'on del
sat\'elite y en la superficie de Tierra y que en estos \'ultimos
se observa como un desplazamiento hacia el violeta. ?`Es esta
diferencia despreciable o es crucial para el funcionamiento del
GPS?. Dado que en un 1 nanosegundo una se\~nal electromagn\'etica
se propaga aproximadamente
  30 cent\'{\i}metros, incluso una diferencia de cientos de nanosegundos crea
dificultades en la exactitud del GPS.

 El reloj del sat\'elite
"corre mas r\'apidamente", i.e., se adelanta unos 45.7
microsegundos por d\'{\i}a comparado con el reloj en la superficie
de Tierra debido, exclusivamente, a efectos gravitatorios de la
posici\'on. Claramente se necesita la Relatividad General para el
funcionamiento correcto del GPS.

\subsection{Efectos de Relatividad Especial debidos al
movimiento \\
(Doppler de segundo orden): \,7,100\,\,ns/d\'{\i}a}

 Adem\'as de los efectos
gravitatorios de posici\'on del sat\'elite y observador en Tierra,
se deben incluir los efectos de Relatividad Especial debidos al
movimiento relativo de los relojes del  sat\'elite y del reloj
fijo en la superficie de la Tierra. El reloj del sat\'elite se
mueve a mayor velocidad que un reloj fijo en  la superficie
terrestre. Sus velocidades son, respectivamente:
 $v_{_{S}}= \,3.874\;{\rm Km}/s$\,;\,
 \,$v _{_{T}}= 465 \;{\rm m}/s$.

La Relatividad Especial nos dice que ``relojes en movimiento
 atrasan" o lo que es lo mismo, a m\'as velocidad corresponde
 un desplazamiento hacia el rojo.

 Por lo tanto, {\it este efecto cinem\'atico Doppler de segundo
 orden ``trabaja en contra" del efecto gravitatorio Einstein}. A una
  altitud de $H= \,3,165 \,{\rm Km}$ se compensan, pero a la altitud
  de los sat\'elites del GPS, $H_{GPS}=\,
 20,183 \,\,{\rm Km}$, el efecto gravitatorio Einstein es mucho
 mayor.

\begin{equation}\label{13}
\frac{dt_{T}}{dt_{S}}=
1-\displaystyle\frac{GM}{r_{_{T}}\,c^2}\,-\,\frac{v^2_{_{T}}}{2\,c^2}
+\displaystyle\frac{GM}{r_{_{S}}\,c^2}\,+\,\frac{v^2_{_{S}}}{2\,c^2}\,.
\end{equation}

El resultado neto final es del orden de $39,000\,\, {\rm ns}$ por
d\'{\i}a.

\subsection{Tiempo coordenado GPS}

N\'otese que {\it para definir las velocidades que aparecen en
(\ref{13}), se est\'a usando el tiempo en el infinito $t$ de la
m\'etrica de Schwarschild y no el tiempo $t_{GPS}$ de relojes en
reposo  en el geoide}. La relaci\'on entre ellos es
\begin{equation}\label{11}
dt_{GPS}= \left(1+2\,\Phi_{0}/c^2 \right)^{\frac{1}{2}}\,dt
\end{equation}
o aproximando a primer orden

\begin{equation}\label{12}
dt_{GPS}=\left(1+\displaystyle\Phi_{0}/c^2 \right)\,dt\,.
\end{equation}

Donde $\Phi_{0}$ es el potencial efectivo de gravedad
  en el geoide, que incluye el t\'ermino centr\'{\i}fugo.
   La correcci\'on negativa,  $\Phi_{0}/c^2$, es del orden de siete
   partes en $10^{10}$.
El tiempo GPS,   $t_{GPS}$, est\'a atrasado con respecto al tiempo
$t$ medido por un sistema inercial en el infinito.

La m\'etrica de Schwarzschild en el sistema de referencia
localmente inercial ECI, en el que son aplicables las ecuaciones
de navegaci\'on y se puede sincronizar, es a orden $O(c^{-2})$:

\begin{equation}\label{13}
ds^2 =  \left(1+\frac{2(V\, -
\,\Phi_{0})}{c^2}\right)\,c^2\,dt_{GPS}^2 - \left
(1-\frac{2\,V}{c^2}\right) (dr ^2 + r^2 d\phi^2)~.
\end{equation}

{\it El resultado neto final, utilizando $t_{GPS}$, es del orden
de $38,600\, {\rm ns}$ por d\'{\i}a, es decir, un error de
$11.580\, {\rm Km}$ por d\'{\i}a.}

Sin embargo las ef\'emerides de los sat\'elites son calculadas en
el sistema ECEF (centrado en la Tierra y que gira con ella) que
para el GPS es el WGS-84(G730).

 Efectuando la transformaci\'on de coordenadas del ECI al ECEF:\\
 $t_{GPS} = t'_{GPS} \,;\,r= r'\,;\, \theta = \theta' \,;\, \phi = \phi' + \omega_{_{T}} \,t'_{GPS}$.

  La m\'etrica de Schwarzschild en el sistema de
referencia ECEF tendr\'a la expresi\'on:
\begin{equation}\label{14}
\begin{array}{cccccccc}
ds^2 &=&\left(1+\displaystyle{\frac{2\,(\Phi\, -
\,\Phi_{0})}{c^2}} \right)c^2\,dt_{GPS}^2 +
2\,\displaystyle{\frac{\omega_{_{T}} r^2}{c^2}}\sin^2\theta\, d\phi'\,dt_{GPS}\\
   & &- \left(1-\displaystyle{\frac{2\,V}{c^2}}\right)
(dr ^2 + r^2 d{\phi'}^2) ~.
\end{array}
\end{equation}

El t\'ermino cruzado es el responsable del efecto Sagnac. Las
coordenadas del sistema de referencia en rotaci\'on, ECEF, no son
globales y son s\'olo v\'alidas hasta el cilindro de luz,
$\omega_{_{T}}r=c$, i.e., hasta un radio $r\approx 28\,\, UA$.

 El tiempo coordenado GPS, $t_{GPS}$, es el tiempo propio de relojes en reposo en
el geoide ($\Phi\,=\,\Phi_ {0}$). Es un tiempo referido al ECEF
(unidades de tiempo de UTC aunque adelantado con respecto a
\'este), pero coincide con un tiempo ficticio de un reloj en
reposo en el ECI, ya que \'este es el \'unico sistema de
referencia en el que se puede establecer la sincronizaci\'on
globalmente.

 La m\'etrica en
el ECI, dada en (\ref{13}), se puede transformar en:

\begin{equation}\label{15}
ds^2 = \left[1+\frac{2(V\, - \,\Phi_{0})}{c^2}\, -
\left(1-\frac{2\,V}{c^2}\right)\,\,
 \frac{dr ^2 + r^2
d\phi^2}{c^2\,dt_{GPS}^2}\right]
 \,c^2\,dt_{GPS}^2.
\end{equation}
La velocidad del sat\'elite en el ECI (supuestas \'orbitas
circulares) es
\begin{equation}\label{16}
 v{_s} = \frac{dr}{dt_{GPS}}  + r\, \frac{d\phi}{dt_{GPS}}\,,
\end{equation}
por lo tanto, a orden $ O(c^{-2})$, el tiempo propio en el
sat\'elite es
\begin{equation}\label{17}
d\tau= ds/c = \left[1+\frac{(V\, - \,\Phi_{0})}{c^2} -
\frac{v{_s}^2}{2c^2}\right] \,dt_{GPS}\,,
\end{equation}
e, integrando sobre una  trayectoria $C$, el tiempo  GPS es
\begin{equation}\label{18}
\int_{C} dt_{GPS}
 = \int_{C} d\tau \,\left[1-\frac{(V\, -
\,\Phi_{0})}{c^2} + \frac{v{_s}^2}{2c^2}\right]\,.
\end{equation}

En esta f\'ormula se observan los cinco  efectos
relativistas principales:\\
1º) En $\Phi_{0}$, tres efectos: monopolo, quadripolo
terrestre y movimiento (potencial centr\'{\i}fugo) de relojes fijos en el ECEF (receptores).\\
 2º) En $V$, el efecto gravitatorio Einstein debido a la
posici\'on del sat\'elite.\\
3ª) En $\frac{v{_s}^2}{2c^2}$, el efecto relativista especial
Doppler de segundo orden debido
al movimiento del sat\'elite.\\
El signo negativo de la suma del segundo y tercer t\'ermino entre
corchetes en la f\'ormula (\ref{18}), significa que el reloj en
\'orbita se adelanta. Para que los relojes at\'omicos de los
sat\'elites aparezcan para el observador en el geoide, ECEF, a la
frecuencia elegida, $10.23 \,{\rm MHz}$, los relojes at\'omicos de
los sat\'elites deben ser ajustados antes de su lanzamiento,
bajando su frecuencia propia a
\[\left[1 - 4.4647 \times 10^{-10} \right]\times 10.23 \,\,{\rm
MHz} = 10.229\, 999\, 995\, 43\,\, {\rm MHz}.\]

Adem\'as hay que corregir el efecto Sagnac y el efecto de
excentricidad (debido a que las \'orbitas no son circunferencias
sino elipses). En el GPS, estos \'ultimos son corregidos por el
receptor.

Finalmente hay que se\~nalar que, al transferir frecuencias,
aparece el efecto Doppler cl\'asico (lineal en velocidad), que en
principio debe ser sustra\'{\i}do por los receptores,  ya que,
dependiendo de la velocidad del receptor,  es $10^{3}-10^{5}$
mayor que los efectos relativistas seculares aqu\'{\i}
considerados. Sin embargo, existen t\'ecnicas de cancelamiento
Doppler por transferencia de frecuencia en dos sentidos que
permiten suprimirle. Una exposici\'on m\'as detallada de los
efectos relativistas en el GPS, puede encontrarse en \cite{ash}.

\section{El proyecto SYPOR para el Galileo}
SYPOR es el anagrama (en franc\'es) de SIstema de POsicionamiento
Relativista. Su Director es B. Coll del Observatoire de Paris. La
idea de base de este proyecto, que se espera sea sometido a un
Programa Marco de la U.E., es la siguiente seg\'un se expone en
\cite {coll}: Una constelaci\'on de sat\'elites, con relojes que
intercambian su tiempo propio entre ellos y con receptores en la
Tierra, es un sistema propiamente relativista. {\it En el SYPOR,
el Segmento de Control est\'a en la constelaci\'on de
sat\'elites}. Por tanto, se invierte el procedimiento usado en los
GNSS newtonianos (GPS, GLONASS). La funci\'on del segmento de
control no es determinar las efem\'erides de los sat\'elites con
respecto a unas coordenadas terrestres, sino determinar las
coordenadas terrestres de los receptores con respecto a la
constelaci\'on de sat\'elites.

Las caracter\'{\i}sticas principales del SYPOR ser\'{\i}an: 1.-
Control externo de {\it todo} el sistema: consta de un dispositivo
en al menos cuatro de los sat\'elites apuntando al ICRS
(International Celestial Reference System). 2.- Control interno de
las {\it partes} del sistema: un dispositivo en cada sat\'elite
que intercambia su tiempo propio con otros sat\'elites. 3.-
Control por el segmento de receptores: consta de un dispositivo en
cada sat\'elite que env\'{\i}a a la Tierra, mediante se\~nales
electromagn\'eticas,
  adem\'as de su tiempo propio, el de sus sat\'elites pr\'oximos.

 En el espacio-tiempo relativista, los frentes de onda
de esas se\~nales, parametrizados  por el tiempo propio de los
relojes, son cuatro familias de conos de luz (hipersuperficies
3-dim) movi\'endose a la velocidad de la luz y realizando un
sistema coordenado  nulo (o luz) covariante. Sistema coordenado
que no existe en el espacio-tiempo newtoniano en el que la luz
viaja a velocidad infinita.

En Relatividad, un sistema de referencia se puede construir a
partir de un sistema de posicionamiento, pero no al rev\'es,
mientras que en la teor\'{\i}a Newtoniana ambos son
intercambiables. En un sistema de referencia, un observador
situado en el origen asigna a los puntos  de su entorno un
conjunto de coordenadas. En un sistema de posicionamiento,
cualquier punto de un entorno arbitrario puede conocer sus
coordenadas propias. Los cuatro tiempos propios de los sat\'elites
 $ (\{\tau_i\};\, i= 1,2,3,4$), le\'{\i}dos por un receptor
constituyen sus coordenadas propias luz o nulas, con respecto a
los cuatro sat\'elites.

La m\'etrica espacio-temporal en coordenadas luz covariantes
  $(\tau_1, \tau_2, \tau_3, \tau_4)$, es sim\'etrica, ver
  \cite{hehl}.
 No hay asimetr\'{\i}a espacio-temporal como en
las coordenadas $(t, e_1, e_2, e_3)$ (una de tiempo coordenado
``$t$'' y tres de espacio ``$e$'') habituales.

 {\it El SYPOR no s\'olo es  un sistema de posicionamiento,
 sino que adem\'as es aut\'onomo}. Definamos lo que
significa  aut\'onomo. Un sistema es aut\'onomo si los receptores,
\'unicamente en base a la informaci\'on recibida durante un
intervalo temporal, pueden conocer
 su trayectoria espacio-temporal as\'{\i} como
la trayectoria  de los cuatro sat\'elites.
  Cuatro sat\'elites emitiendo, sin
necesidad de sincronizaci\'on,  no s\'olo su tiempo propio
$\tau_i$, sino tambi\'en  los tiempos propios $\tau_{ij}$ de los
tres sat\'elites pr\'oximos recibidos  por el sat\'elite $i$ en
$\tau_i$ ($ \{\tau_i, \tau_{ij}\};\, i \neq j; \,i, j = 1,2,3,4$),
constituyen un sistema aut\'onomo de posicionamiento. Los
diecis\'eis datos $ \{\tau_i, \tau_{ij}\}$ recibidos por un
usuario, constituyen una carta local del sistema coordenado nulo
covariante, del atlas obtenido mediante toda la constelaci\'on de
sat\'elites. Si la constelaci\'on est\'a situada alrededor de la
Tierra, como en el Galileo, se necesitar\'{\i}an m\'as de cuatro
sat\'elites y las redundancias ser\'{\i}an usadas para modelar el
propio campo gravitatorio terrestre.

La aplicaci\'on del proyecto SYPOR  al Galileo, har\'{\i}a  de
\'este un sistema de posicionamiento aut\'onomo plenamente {\it
relativista}. No habr\'{\i}a que corregir los errores  a que dan
lugar los efectos relativistas (se obtiene la suma de toda la
serie postnewtoniana en $c^{-n}$) y tendr\'{\i}a ventajas
tecnol\'ogicas adicionales.

\section*{Acknowledgments}
Deseo agradecer a Bartolom\'e Coll, las ideas que me ha comunicado
sobre el proyecto SYPOR y a Angel  San Miguel y Francisco Vicente,
una lectura cr\'{\i}tica del manuscrito. Este trabajo ha sido
realizado en parte, gracias a la ayuda de la Junta de Castilla y
Le\'on al projecto VA014/02.

\end{document}